\input{psfig.sty} 

\documentclass{emulateapj}

\def\saxj{SAX J1808.4--3658 }  
\def\msun{{\rm M}_\odot} 
\def\Porb{P_{\rm orb}}
\def\Pdot{\dot{P}_{\rm orb}} 
\def\Pspin{P_{\rm spin}} 
 
\def\T0{T^*_0}
\def\asini{a_1 \sin i}

\begin{document} 
 
\title{Revised orbital parameters of the accreting millisecond pulsar \saxj} 
 
\author{A. Papitto\altaffilmark{1}, M.T. Menna\altaffilmark{1},
L. Burderi\altaffilmark{1}, T. Di Salvo\altaffilmark{2}, F. D' Antona
\altaffilmark{1}, N. R. Robba\altaffilmark{2}}
 
\altaffiltext{1}{Osservatorio Astronomico di Roma, via Frascati 33, 
00040 Monteporzio Catone (Roma), Italy email: papitto@mporzio.astro.it,  
burderi@mporzio.astro.it} 
\altaffiltext{2}{Dipartimento di Scienze Fisiche ed Astronomiche, 
Universit\`a di Palermo, via Archirafi 36 - 90123 Palermo, Italy} 
 
\begin{abstract} 
 
We present temporal analysis of the three outbursts of the X-ray 
millisecond pulsar SAX J1808.4--3658 that occurred in 1998, 2000 and 
2002.  With a technique that uses the $\chi^2$ obtained with an epoch 
folding search to discriminate between different possible orbital 
solutions, we find an unique solution valid over the whole 
five years period for which high temporal resolution data are 
available. We revise the estimate of the orbital period, 
$P_{\rm orb}=7249.1569(1)$ s and reduce the corresponding error
by one order of magnitude with respect to that previously 
reported. Moreover we report the first constraint on the orbital period 
derivative, $-6.6 \times 10^{-12} < \Pdot < +0.8 \times 10^{-12}$  
s s$^{-1}$.  These  values allow  us to 
produce, {\it via} a folding technique, pulse profiles at any given 
time. Analysis of these profiles shows that the pulse shape is 
clearly asymmetric in 2002, occasionally showing a secondary 
peak at about $145^{\circ}$ from the main pulse, which is quite 
different from the almost sinusoidal shape reported at the beginning 
of the 1998 outburst.   
 
 
\end{abstract} 
 
\maketitle 
 
\section{Introduction} 
The X-ray transient SAX J1808.4-3658 was discovered in September 1996 
when it exhibited an outburst detected by the BeppoSAX Wide Field 
Cameras (in 't Zand et al. 1998).  The source showed type-I 
X-ray bursts that led to the identification of the compact object as a 
neutron star (hereafter NS) and to the derivation of a distance of 
about $2.5$ kpc (in 't Zand et al. 2001). 
The source was found in 
outburst again in April 1998, when the high temporal resolution of the 
proportional counter array (PCA) on-board the Rossi X-ray Timing 
Explorer (RXTE), allowed the discovery of coherent $401$ Hz pulsations, 
making this source the first known accretion-driven 
millisecond pulsar (Wijnands \& van der Klis 1998). 
Timing analysis performed on data collected 
over the period $11-18$ April 1998, allowed the determination of the 
orbital parameters reported in Table \ref{tab1} (Chakrabarty \& Morgan 1998, 
CM98 hereafter).  
SAX J1808.4-3658 was again detected in outburst (and observed with RXTE/PCA) 
in January 2000 (van der Klis et al. 2000) and in October 2002 (see 
Wijnands 2004 for a review).

Correct ephemerides are of capital importance for the study of this
and similar systems, since these give us the possibility to 
monitor the evolution of the spin and the orbital period of the system. 
These, in turn, give important information on the mechanism of
the acceleration of a NS to millisecond periods and on the
evolutionary history of the system, respectively. Orbital ephemerides
are also important for performing studies, for instance, in other 
wavelengths. In particular any search for radio pulsations during 
quiescence (to test the hypothesis that the radio pulsar switches 
on during quiescence) needs correct orbital parameters to perform 
coherent search for pulsations and to improve by several orders of
magnitude the detection sensitivity. 
In this paper we apply a temporal analysis on the three most recent  
outbursts of SAX J1808.4--3658 in order to improve the orbital parameters
of the system; we find a new orbital solution which appears to be valid
from 1998 to 2002 (the entire period of the RXTE observations of \saxj
up to date). 
In \S 2 we present the timing technique we developed to
obtain the revised estimate of the orbital period; in \S 3 we apply this
technique to the available RXTE/PCA datasets of the 1998, 2000, and 2002
outbursts of the source; in \S 4 we draw our conclusions.

\section{A method of timing for the orbital period} 
Let us consider a binary system of ``true'' orbital period ${\cal
P}_{\rm orb}$, orbital separation ${\cal A}$, and negligible
eccentricity in which a source (hereafter the primary) orbits the
barycenter of the system at a radius ${\cal A}_1 = {\cal A} \,
m_2/(m_1 + m_2)$ (where $m_1 = M_1/\msun$ and $m_2 = M_2/\msun$ are
the masses of the primary and its companion, the secondary,
respectively).  The orbital motion causes delays in the times of
arrival, $t_{\rm arr}$, of a signal -- emitted by the primary at the
time $t_{\rm em}$ -- on the plane passing through the barycenter and
perpendicular to the line of sight: $t_{\rm arr} = t_{\rm em} +
Z(t_{\rm em})/c$, where $Z(t) = {\cal A}_1 \sin i
\sin\left[(2\pi/ {\cal P}_{\rm orb} )(t-{\cal T}^{*})\right]$ 
is the distance of the primary from the plane defined above, $i$ is
the inclination angle between the normal to the orbital plane and the
line of sight, ${\cal T}^{*}$ is the time of passage at the ascending
node, and $c$ is the speed of light.  With an observational estimate of 
the orbital parameters $a_1 \, \sin i$, $\Porb$, and $T^{*}$, we can 
correct the arrival times of photons emitted by the primary for the 
delays induced by the orbital motion.  A first order correction gives:
\begin{equation} 
\label{eq:corr} 
t_{\rm em} = t_{\rm arr} - \frac{1}{c} 
a_1 \sin i \sin\left[ \frac{2\pi}{\Porb} 
\left(t_{\rm arr} - T^* \right)\right]  
\end{equation}

The high temporal resolution RXTE data of \saxj span a period of
$\Delta T_{\rm data}\sim 4.5$ yr, from April 1998 to October 2002,
during which \saxj performed $N_{\rm max}\sim 19700$ orbital cycles
and $n_{\rm max} \sim 57$ billion spin cycles.  In order to perform a
timing analysis over the whole $\Delta T_{\rm data}$ we must
unambiguously associate $n$, i.e. the number of elapsed spin cycles
since the 1998 outburst, to any given $t_{\rm arr}$. 
Neglecting any error on the spin period estimate $P_{\rm spin}$,
induced by the errors in the orbital correction, this association
would be possible if the intrinsic error on the estimate of $P_{\rm
spin}$, $\sigma_{\rm Pspin}$, is such that $ \sigma_{\rm Pspin}
\la \Pspin / n_{\rm max} \sim 4 \times 10^{-14} \; {\rm s.}$ 
CM98  reported  $\sigma_{\rm Pspin}  \sim 5
\times  10^{-12}$  s, thus ruling  out   the possibility  of making an
overall timing analysis over $\Delta T_{\rm data}$.

However, the large value of $N_{\rm max}$ suggested us to develop a
method of timing of the orbital period $\Porb$ in order to improve the
orbital period estimate of CM98. Our idea is to estimate the time of
passage at a given point in the orbit, e.g. the ascending node
$T^{*}$, for any given orbital cycle $N$, and to fit these times $T^{*}$s
{\it vs} the corresponding $N$ to improve the estimate of $\Porb$ and
to give a value (or an upper limit) for the orbital period derivative
$\Pdot$. This timing can be done providing that: i) we can
unambiguously associate $N$, i.e. the number of elapsed orbital cycles
since the 1998 outburst, to any given $T^*$. This is possible if
the error on $\Porb$, $\sigma_{\rm Porb}$, is such that $\sigma_{\rm
Porb} \la \Porb / N_{\rm max} \sim 0.4
\; {\rm s}$. CM98  reported  $\sigma_{\rm Porb} \sim 1 \times 10^{-3}$ 
s, thus making this association possible; ii) we are able to
estimate the $T^{*}$s for any orbital cycle $N$ with an error {\it
smaller} than $\sigma_{T_{\rm pred}^*} = ( \sigma_{\T0}^2 + N^2
\sigma_{\rm Porb}^2 )^{1/2} \sim N \sigma_{\rm Porb} $ obtained from
the propagation of the CM98 errors in the ``theoretical'' computation
of the $T_{\rm pred}^*$ at a given $N$ as $T_{\rm pred}^*(N) = \T0 + N
\times \Porb$, where $\T0$ is the time of passage at the ascending node
at the beginning of the 1998 outburst, estimated by CM98.

Our strategy to estimate the values of $T^{*}$s relies on the fact
that the correction of the photon arrival times with
eq.\ (\ref{eq:corr}) induces a ``smearing'' in a pulse profile obtained
by an epoch folding search performed on the corrected lightcurve
because the corrections performed are affected by the errors in the
estimate of the orbital parameters. Indeed, simple differentiation of the
Doppler effect formula $\Pspin(t) = \Pspin \times (1 - v(t)/c)^{-1}$, where
$v(t)= \dot{Z}(t)$ is the speed along the line of sight, demonstrates
that the leading term in the error induced by eq.\ (\ref{eq:corr}) on
$\Pspin$ is $\sigma_{\rm Pspin} \la | \sin \phi | \Pspin \asini (2\pi/
\Porb)^2 N_{\rm max}
\sigma_{\Porb}$, where $\phi = (2\pi/\Porb) (t - T^*)$ is the orbital phase.
Since $ | \sin \phi | \le 1$ we have
$\sigma_{\rm Pspin} \la \Pspin \asini (2\pi/ \Porb)^2 N_{\rm max}
\sigma_{\Porb}$. This means that folding a lightcurve of length
$\sim 1$ hour $\sim \Porb/2$ at a period close to $\Pspin$ results into
delays up to $D_{\rm max} \sim \pi \asini (2\pi/ \Porb) N_{\rm max}
\sigma_{\Porb}$.
Adopting the CM98 estimates of the orbital parameters
and the related errors we get $D_{\rm max} \sim 3.2 \times
10^{-3}\,{\rm s} \sim 1.3 \, \Pspin$ for the 2002 data, 
which means that a significant smearing is
is induced by $\sigma_{T_{\rm pred}^*}
\sim N_{\rm max} \sigma_{\rm Porb}$, which represents our ignorance of 
the ``true'' time of ascending node passage in the 2002 data determined by 
the original error on $\Porb$ propagated through the factor 
$N_{\rm max}$.

Therefore we can ``experimentally'' estimate the $T^{*}$s in the
2000 and 2002 outbursts by correcting each $\sim 1$ hour length observation
with eq.\ (\ref{eq:corr}) in which several trial values of $T^{*}$ are adopted.
We then choose the $T^{*}$ for which the corresponding corrected 
lightcurve gives a folded pulse profile with minimum
smearing, obtaining an error in the determination of that particular
$T^{*}$ that is expected to be {\it smaller} than the corresponding
$\sigma_{T_{\rm pred}^*}$\footnote{Note that, to estimate $T^*$ 
selecting the lightcurve with the minimum smearing, the $\sigma_{\rm Pspin}$ 
(caused by the uncertainty in the knowledge of the true time of ascending 
node passage) must induce a genuine broadening of the folded pulse profile,
and not delays more or less constant during each
observation that can be compensated, in the folding search procedure, by
the selection of a different spin period.
Since the induced delays are not constant, but vary with time as $\sin \phi$, 
the epoch folding search procedure must be performed 
over a time span that is a relevant fraction of the orbital period,
so that $\sin \phi$ varies significantly. This condition is verified
by our sample of observations that last $\sim 1$ hour $\sim \Porb/2$.}.

Since a folded pulse profile obtained by folding a lightcurve corrected 
with the ``true'' time of ascending node passage (and therefore not smeared) 
has a bigger pulsed fraction than one obtained by folding a lightcurve 
corrected with a ``wrong'' time of ascending node passage (and therefore 
significantly smeared), the experimental estimate of each
$T^{*}$ can be done by the following procedure: a) for a given
observation we produce a number of corrected lightcurves applying
eq.\ (\ref{eq:corr}) with different values of the time of ascending
node passage $T^{*}_{\rm trial}$; b) we then perform on each lightcurve an 
epoch folding search for periodicities fitting the $\chi^2$ {\it vs} spin 
period curve close to its 
maximum with a Gaussian profile, to derive the maximum $\chi^2$ value,
$\chi^2_{\rm max}$, for the given value of $T^{*}_{\rm trial}$; 
c) we plot these $\chi^2_{\rm max}$ {\it vs} their corresponding 
$T^{*}_{\rm trial}$ and we fit the part of this curve close to its maximum
again with a Gaussian choosing, as our experimental estimate of the time of
passage at the ascending node $T^{*}$, the value of $T^{*}_{\rm trial}$ 
for which this Gaussian has a maximum.  The uncertainties on $T^{*}$ 
are computed using post-fit residuals, i.e.\ the average fluctuation of the 
$T^{*}$ values vs.\ the orbital cycle number $N$ with respect the average
trend (see below). 
This is the more conservative way of estimating the errors in the case of
data with similar statistical significance as expected since we have used
datasets of comparable length (see Tab. 1): indeed the post-fit errors are  
$\sim 8$ times bigger than the errors based on a fit of the top of the Gaussian.

This technique supplies us with values of the epoch of ascending node
passage $T_N^*$ that can be compared with the ones predicted using the
original estimate of orbital parameters $T_{\rm pred}^*(N) = \T0 + N
\times \Porb$. We therefore plot the delays $\Delta T_N^{*} = T_N^{*}
- T_{\rm pred}^{*}(N) $ versus $N$ (we note that
$N$ is derived  unambiguously as the integer part of 
$ (T_N^{*} - \T0)/ \Porb$ under the assumption that $| T_N^{*} - \T0 | 
< \Porb$ that we have also verified {\it a posteriori}),
fit the data with the relation 
\begin{equation}
\label{eq:fit}
\Delta T_N^*= \alpha + \beta N + \gamma N^2 
\end{equation}
and finally obtain our improved orbital parameters as
${\bf \T0} = \T0 + \alpha$, $\sigma_{\bf \T0} = \sigma_{\alpha}$,
${\bf \Porb} = \Porb + \beta$, $\sigma_{\bf \Porb} = \sigma_{\beta}$,
$\Pdot = 2\gamma /{\bf \Porb}$, $\sigma_{\Pdot} = (2/{\bf \Porb})
[ (\gamma /{\bf \Porb})^2 \sigma_{\beta}^2 + \sigma_{\gamma}^2 ]^{1/2}$. 
Although we devoleped this technique independently, we subsequentely found
that a similar one had already been used by Paul et al. (2001) and Naik \& Paul
(2004).

\section{Observations and analysis} 
Throughout this paper we used the ToO public domain data of the PCA
on-board RXTE (Bradt et al. 1993). In particular, we analyzed
data from the PCA (Jahoda et al.\ 1996), which is composed by a set of
five Xenon proportional counters operating in the 2--60 keV energy
range with a total effective area of $\sim 7000$ cm$^2$. 
We considered a sample (listed in Table \ref{tab2}) of the
available PCA observations of SAX J1808.4--3658 taken during the 1998,
2000 and 2002 outbursts in which the spin modulation was particularly 
strong.
For the temporal analysis we used event mode data with
64 energy channels and $122\mu$s temporal resolution; the arrival
times of all the events were reported to the solar system barycenter.
 
For the first 2002 observation (see Table \ref{tab2}) we produced a set of
lightcurves performing orbital correction adopting eq.\ (\ref{eq:corr}),
with $\asini$ and $\Porb$ as determined by CM98. According to the
method described above, we adopted different $T^*$ spanning a range around
the predicted value resulting from $T_{\rm pred}^*(N_{\rm G}) = \T0 +
N_{\rm G} \Porb$, where $N_{\rm G}$ is our initial guess of $N$:
$N_{\rm G} = {\rm INT}
\{(t_{02} - \T0)/\Porb\}$, with $t_{02}$ the starting time of the first 2002 
observation. The $T^*$ range spans $\sim 40$ s in steps of 2 s as
$\sigma_{T_{\rm pred}^*} \la N_{\rm max} \sigma_{\rm Porb} \sim 20$ s
adopting the CM98 estimates.
We then performed an epoch folding search on each corrected lightcurve, 
but, surprisingly, no pulsation was detected in any of them.  

Thus we extended the range of trial values for $T^*$ to explore the entire
orbit until we obtained lightcurves for which the epoch folding search
produced $\chi^2$ peaks (typically with maxima of $\sim 3000$ over an average
value of $\sim 40$) around periods close to $\Pspin$.
Following the method described above we then fitted the curve of the 
maxima of the $\chi^2$ (obtained from the epoch folding search on each
lightcurve) {\it vs} the corresponding $T^*$ with a Gaussian and derived 
our best estimate of the time of ascending node passage for the first 2002
observation $T^*_{N_{02}}$. We then estimated $N_{02}$, i.e. the number 
of orbital cycles elapsed since $\T0$ as 
$N_{02} = INT\{ (T^*_{N_{02}} - \T0)/\Porb \} = 19657$ 
and therefore adopted a new estimate of the orbital period 
$P_{\rm orb}^{\rm new}$ that is obtained from $P_{\rm orb}^{\rm new} = 
(T^*_{N_{02}} - \T0)/N_{02}$. 
We checked the consistency of our provisional estimate of the orbital 
period $P_{\rm orb}^{\rm new}$ demonstrating that $N_{02}$ can be 
unambiguously associated  to $T^*_{N_{02}}$ by verifying that 
$INT\{ (T^*_{N_{02}} - \T0)/\Porb \} = INT\{ (T^*_{N_{02}} - 
\T0)/P_{\rm orb}^{\rm new} \}$. 

With this new estimate we repeated the procedure described above on
each observation listed in Table \ref{tab2}, with different $T^*$ spanning a
range of 40 s around the new predicted value resulting from $T_{\rm
new}^*(N_{\rm G}) = \T0 + N_{\rm G} P_{\rm orb}^{\rm new}$, where
$N_{\rm G} = {\rm INT} \{(t - \T0)/\Porb\}$, with $t$ the starting
time of each observation. Encouragingly we always detected the
pulsation, although with different strength (as deduced from the
different maximum $\chi^2$), by performing the epoch folding search on the
corrected lightcurves.  This allowed us to determine a set of 15
$T^*_N$, one for each observation.

In Fig. 1 (top and middle panel) we plot the delays between the
$T_N^{*}$ values and the values predicted adopting the CM98 estimates
$\Delta T_N^{*} = T_N^{*} - T_{\rm pred}^*(N)$ versus the number of
orbital cycles elapsed since $\T0$: where $N = {\rm INT}
\{(T_N^{*} - \T0)/\Porb\}$, and $T_{\rm pred}^*(N) = \T0 + N \times
\Porb$.  It should be noted that all the delays are $\le \Porb$ which
demonstrates {\it a posteriori} the consistency of our method, since
we were able to associate unambiguously $N$ to any $T_N^{*}$.  Fitting
these delays with eq.\ (\ref{eq:fit}) we find the improved orbital
parameters and the first constraint on the orbital period derivative 
that are reported in Table \ref{tab1}.

We finally corrected all the observations using eq.\ (\ref{eq:corr})
with the improved value of the orbital period ${\bf \Porb}$ 
and again we produced, for each observation, a set of lightcurves with 
different $T^*$ spanning a range around the predicted value resulting 
from ${\bf T}_{\rm pred}^*(N) = {\bf \T0} + {\bf \Porb} \times N$ 
that allowed us to determine the new set of
15 ${\bf T}^*_N$.  We finally computed the ``post fit'' residuals as
${\cal R}(N) = {\bf T}_N^{*} - {\bf T}_{\rm pred}^*(N)$ in order to
check if the residuals were now compatible with zero.  In Fig. 1
(bottom panel) we plotted ${\cal R}(N)$ versus $N$ to demonstrate the
consistency of our method.
   
\section{Discussion} 
For the estimate of the $T^*$s we have corrected the photons arrival times
with eq.\ (\ref{eq:corr}), adopting the estimate of $\Porb$ and
$ x = \asini/c$ given by CM98.  This seemed appropriate since we have
demonstrated that, adopting the value of $\sigma_{\Porb}$ and $\sigma_{x}$ 
reported by CM98, the leading term in $\sigma_{\rm Pspin}$ depends only on
$\sigma_{T_{\rm pred}^*} \sim N_{\rm max} \sigma_{\rm Porb}$.
However, at the end of our analysis, we found that our best-fit orbital period 
is incompatible with that reported in CM98, and this explains why the
use of the ephemeris published in CM98 allows to recover the coherent 
pulsations only during the 1998 outburst.
We want to stress that a test of the correctness of our orbital
solution is given by the fact that with our revised orbital
parameters we can recover the pulsations in all the observations between
1998 and 2002.

We also found a $\Pdot \neq 0$ (at $1 \sigma$ confidence level) which 
means that in principle its effects on orbital parameters
during $\Delta T_{\rm data} \sim 4.5$ yr must be discussed.
Indeed the orbital period has varied and, because of the three
outburst that occurred during $\Delta T_{\rm data}$, $m_1$ and $m_2$
varied as a consequence of mass transfer.  Both these variations
produce, because of third Kepler's law, a variation of the orbital
separation along $\Delta T_{\rm data}$ which reflects in a
secular variation of $x$.  We therefore checked {\it a posteriori}
that both the secular variation of $x$ and $\Porb$ did not affect our
determination of the orbital parameters of the system.
By differentiating ${\cal P}_{\rm orb}(t) = {\cal P}_{\rm orb}(t=t_0)+
\Pdot \times (t-t_0)$ and the third Kepler's law
(assuming that the inclination angle has remained constant),  
we have demonstrated that the uncertainties on $\Porb$
and $x$ in the 2002 outburst are always dominated by the uncertainties
in their original determination as given by CM98. In particular we 
obtain that the uncertainty on $x$ induced by mass transfer is of the
order of $10^{-9} d_{2.5}^2 m_1^{-1} R_6$ times the uncertainty reported 
by CM98, as soon as the orbital period derivative is 
$|\Pdot| \le 3 \times 10^{-12}$, the number of orbital cycles elapsed 
since the 1998 outburst is $N \la N_{\rm max}$, and the averaged mass loss
rate by the secondary during the whole period $\Delta T_{\rm data}$ is
not much greater than the average mass transfer rate during the three
outbursts, which is $\dot{M}_{\rm outb}/(10^{-10} 
\msun {\rm yr}^{-1}) = 3.7\, d_{2.5}^{2} m_1^{-1} R_6$ (as can be easily 
determined from an analysis of the X-ray luminosity during the 1998,
2000, and 2002 outbursts), where $d_{2.5}$ is the source distance in units of
2.5 kpc and $R_6$ is the NS radius in units of $10^6$ cm.
Therefore, our assumption of using $\asini$ and $\Porb$ as given by
CM98 in eq.\ (\ref{eq:corr}) is justified {\it a posteriori},
since their associated uncertainties, even taking into account the
effects of orbital evolution, are always comparable with those
estimated by CM98.

The improved estimate of $\Porb$ derived in this paper (whose associated
error, $\sigma_{\Porb}$, is 10 times smaller than that reported by CM98)
allowed us to correct a lightcurve using eq.\ (\ref{eq:corr}) with the
improved value of the orbital period ${\bf \Porb}$
and the predicted value of $T^*$ resulting from ${\bf
T}_{\rm pred}^*(N) = {\bf \T0} + {\bf \Porb} \times N$. 
As discussed in the fourth paragraph of \S 2, folding this lightcurve modulo
$\Pspin$ for any time interval $\Delta T_{\rm fold} \ge \Porb/2$ introduces
delays in the reconstruction of the pulse profile up to $D_{\rm max} \sim 2
\asini (2\pi/ \Porb) N_{\rm max}
\sigma_{\Porb}$ (since $|\sin \phi|\le 1$ is a
periodic function of period $\Porb/2$ and average value $2/\pi \sim 0.6$).
Adopting our new estimate of $\sigma_{\Porb}$
we get $D_{\rm max} \sim 2.0 \times
10^{-4}\,{\rm s} \sim 0.08 \, \Pspin$ for the 2002 data.  
Which means that if the number of bins in the pulse profile is up to few tens,
we do not expect a significant distortion because of the uncertainties
in the orbital parameters. 
In Fig. \ref{fig2} we compare two pulse profiles obtained by folding 
modulo $\Pspin$ (the CM98 estimate) $\sim 20$ and $\sim 8$ hours of data 
of the 1998 and 2002 outbursts, respectively,
corrected with eq.\ (\ref{eq:corr}). It is evident that the 2002 
pulse profile has significantly varied with respect to the almost sinusoidal 
shape showed in the 1998 outburst; in particular we find a secondary peak
at about $145^{\circ}$ from the main pulse. This variability of the pulse 
profile suggests that strong caution should be taken in carrying out the 
standard pulse timing analysis when 
computing the delays of the time of arrival of a pulse by cross-correlating
the locally folded pulse profile with a sinusoid, as done in CM98. 
Temporary variations of the pulse shape could introduce spurious
delays that may significantly affect the robustness of some derived
parameters e.g.\ the local spin period derivative. 

In our analysis we found 
$-6.6 \times 10^{-12} < \Pdot < +0.8 \times 10^{-12}$  s s$^{-1}$
(90\% c.l.).
Indeed, orbital evolution calculations show that the magnitude of 
the $\Pdot$ caused by angular momentum
losses associated with the emission of gravitational radiation is 
expected to be one order of magnitude
smaller, $\Pdot \sim - 3.0 \times 10^{-13} (n-1/3) m_1 (m_1+m_2)^{-1/3}
P_{\rm 2h}^{-2/3}$ s s$^{-1}$, where we assumed a mass--radius relation
for the secondary $R_2 \propto m_2^n$, and $P_{\rm 2h}$ is the orbital
period in units of 2 hours. However the error in the estimate of
$\Pdot$ scales as $\Delta T_{\rm data}^2$. This means that if, as
expected, the next outburst of \saxj will occur in 2005,
we will reduce $\sigma_{\Pdot}$ by a factor 
$\sim (4.5\,{\rm yr}/7 \,{\rm yr})^2 \sim 0.4$.
This new measure could severely constraint some of the evolutionary
scenario proposed: for instance, the degenerate nature of the companion 
(that has been proposed to be a brown dwarf by Bildsten \& Chakrabarty 2001) 
will be effectively probed, since $n\le 0$, as expected for a degenerate
companion, would imply $\Pdot > 0$.

\acknowledgements 
This work was partially supported by the MIUR.


\begin{deluxetable}{llll} 
\tabletypesize{\scriptsize}
\tablewidth{8.5cm}
\tablecaption{RXTE observations of SAX J1808.4-3658 used in this work \label{tab2}} 
\tablehead{\colhead{Observation ID} &\colhead{Time of observation} 
&\colhead{Phase\tablenotemark{1}} &\colhead{Phase(CM)\tablenotemark{2}}  }   
\startdata 
\cutinhead{1998 outburst}
30411-01-01-02S & 50914  19:26 - 19:52 &0.18799(8)&0.18799(8)  \\ 
30411-01-03-00  & 50919  17:21 - 18:25 &0.74694(8) &0.74724(8)  \\ 
30411-01-05-00  & 50921  07:48 - 08:57 &0.84090(8) &0.84130(8)  \\
\cutinhead{2000 outburst}		                       
40035-05-03-00  & 51576  02:30 - 03:29 &0.90879(8) &0.94999(8)  \\ 
40035-05-05-00  & 51582  00:36 - 01:29 &0.47181(8) &0.51338(8)  \\
\cutinhead{2002 outburst}		                       
70080-01-01-020 & 52564  03:59 - 04:46 &0.24819(8) &0.35096(8)  \\ 
70080-01-01-020 & 52564  08:43 - 09:41 &0.59680(8) &0.69959(8)  \\ 
70080-01-02-000 & 52565  08:29 - 09:27 &0.37313(8) &0.47597(8)  \\ 
70080-01-02-000 & 52565  10:07 - 10:58 &0.21384(8) &0.31669(8)  \\ 
70080-01-02-04  & 52566  03:24 - 04:21 &0.79477(8) &0.89766(8)  \\ 
70080-01-02-06  & 52567  03:07 - 04:04 &0.56876(8) &0.67172(8)  \\ 
70080-01-02-08  & 52568  02:50 - 03:47 &0.35035(8) &0.45337(8)  \\ 
70080-01-02-14  & 52571  01:57 - 02:55 &0.67047(8) &0.77367(8)  \\ 
70080-01-03-02  & 52573  23:30 - 00:27 &0.20678(8) &0.31017(8)  \\ 
70080-02-01-04  & 52575  22:55 - 23:52 &0.75258(8) &0.85608(8) 
\enddata
\tablenotetext{1}{Orbital phase of the beginning of each observation, referred to 
$T_0^*=50914.878465(1)MJD$ and derived using our estimate of $P_{orb}=7249.1569(1)s$}
\tablenotetext{2}{Orbital phase of the beginning of each observation, referred to 
$T_0^*=50914.878465(1)MJD$ and derived using CM98 estimate of $P_{orb}=7249.119(1)s$}

\end{deluxetable}


\begin{deluxetable}{lrr} 
\tabletypesize{\scriptsize}
\tablewidth{8.5cm}
\tablecaption{Improved orbital parameters of SAX J1808.4-3658 \label{tab1}} 
\tablehead{\colhead{} &\colhead{CM98} &\colhead{This work}} 
\startdata 
$\Porb$ (s)                    & $7249.119(1)$ & $7249.1569(1)$     \\
$\Pdot \,(10^{-12}$ s s$^{-1})$  &	
& $-6.6  < \Pdot < +0.8 $ 					    \\
${\T0}$ (MJD)\tablenotemark{a} & $50914.878465(1)$  & $50914.878468(4)$  \\
\enddata
\tablecomments{Numbers in parentheses are the 1 $\sigma$ uncertainties
in the last significant figure. Limits on $\Pdot$ are quoted at 90\% 
confidence level.}
\tablenotetext{a}{CM98 reported a value of $50914.899440(1)$ MJD 
for the epoch of superior conjunction, i.e. when the NS is behind the 
companion; as, in this work, we considered the epoch of ascending node passage
as a reference time, the CM98 reference time reported here has been 
decremented by $\Porb/4$.}
\end{deluxetable}


\begin{figure}
\epsscale{0.8}
\plotone{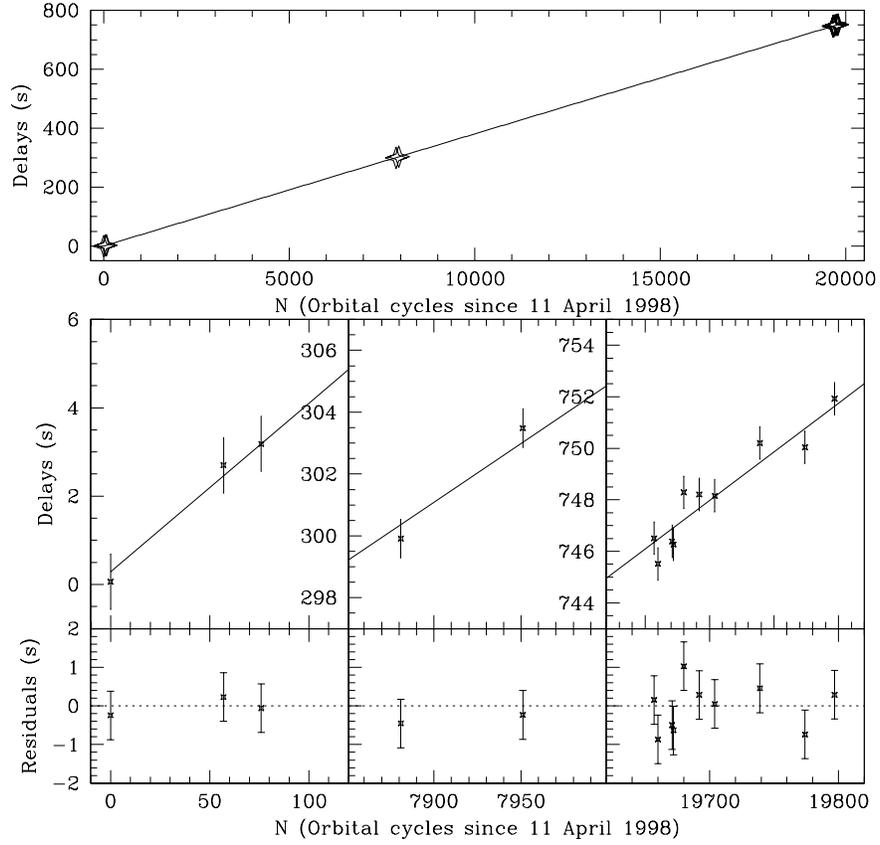}
\caption{Upper panel: Delays, $\Delta T_N^{*}$, with respect to the ephemeris 
of CM98 vs the number of orbital cycles $N$ elapsed since April 11, 1998. 
Note that marks are much larger then error bars. 
Middle panel: expanded view of the three regions where points are clustered. 
Bottom panel: residuals ${\cal R}(N)$ vs $N$.  \label{fig1}}
\end{figure}


\begin{figure}
\epsscale{0.6}
\plotone{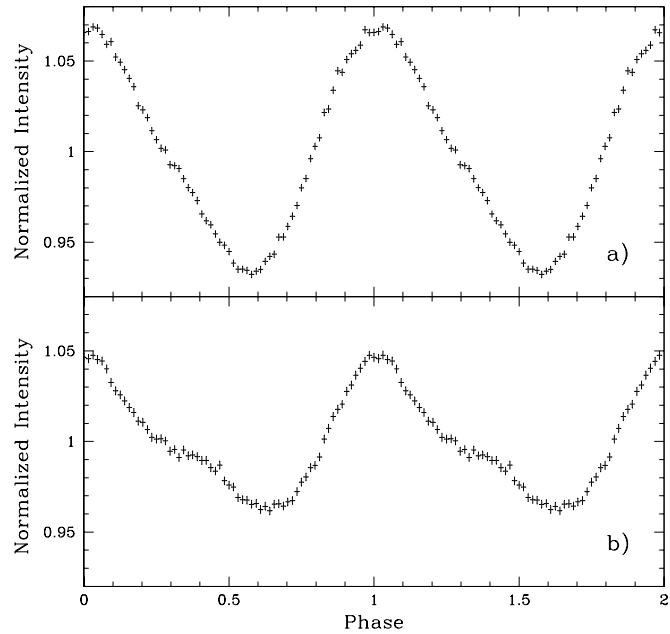}
\caption{64-bin folded pulse profiles ($\Pspin = 2.4939197575$ ms, 
as reported by CM98). 
Upper panel a) refers to a folded light curve from 18-04-1998 03:12:01
to 19-04-1998 00:57:43. Lower panel b) refers to a folded light curve
from 17-10-2002 03:59:35 to 17-10-2002 11:59:35. \label{fig2}}
\end{figure}

\end{document}